\documentstyle[epsfig]{aipproc}

\begin{document}
\title{Studies of s-channel Resonances\\ 
at the CLIC Multi-TeV $e^+e^-$ Collider}

\author{Marco~Battaglia$^*$, Stefania~De~Curtis$^{**}$, 
Daniele~Dominici$^{**}$, Arnaud~Ferrari$^{*}$ and Jari~Heikkinen$^{***}$}
\address{$^*$CERN, CH-1211 Geneva 23 Switzerland\\
$^{**}$INFN and Universita' di Firenze, I-50125 Firenze Italy\\
$^{***}$Helsinki Institute of Physics, FIN-00014 Helsinki Finland}
\maketitle

\vspace*{-0.75cm}
\begin{abstract}
Several models predict the existence of new resonances in the multi-TeV
region, which should be accessible in $e^+e^-$ collisions by s-channel
production. In this paper, we review the phenomenology of some specific
models and we present a preliminary study of the potential of the future
CLIC collider for the determination of their nature and properties.
\end{abstract}

\section{Introduction \label{sec:intro}}

While the core of the physics program of a 500~GeV linear collider can  
already be defined on the basis of the present data, signals from new physics
that can be probed by a collider such as {\sc Clic}~\cite{clic} at 
1~TeV$<\sqrt{s}<$~5~TeV, belong to a significantly broader domain. Still, the 
most striking manifestation of new physics in the multi-TeV region will come 
from the sudden increase of the $e^+e^- \rightarrow f \bar f$ cross section 
indicating the s-channel production of a new particle. There are several 
theories that predict the existence of such a resonance. A first class consists
of models with extra gauge bosons such as a new neutral $Z'$ gauge boson. 
This is common to 
both GUT-inspired $E_6$ models and to Left-Right symmetric models.
They are discussed in Section~II. Additional resonances are also 
predicted by recent theories of gravity with extra 
dimensions in the form of Kaluza-Klein graviton and gauge boson excitations 
~\cite{rs}. Models of dynamical electroweak symmetry breaking also predict the 
existence of new resonances in the TeV region. In particular, we shall consider
in Section~III the degenerate BESS model, which predicts a pair of narrow and 
nearly degenerate vector and an axial-vector states~\cite{dbess}.
The experimental study of such resonances at a multi-TeV collider will have 
to accurately measure their masses, widths, production and decay properties to
determine their nature and identify which kind of new physics has been 
manifested. 

\section{Study of a $Z'$ Boson at CLIC}

One of the simplest extensions of the SM is to introduce an additional 
$U(1)$  gauge symmetry, whose breaking scale is close to the Fermi scale. 
This extra symmetry is predicted in some grand unified theories and in 
other models. The extra $Z'$ associated to this symmetry naturally mixes 
with the SM $Z$ but the mixing angle is strongly constrained by precision 
electroweak data to be of the order of few mrad. Furthermore, direct searches 
for a new $Z'$ boson set a lower mass limit around 600~GeV. As a reference, 
an extra $Z'$ boson having the same couplings as the SM $Z^0$ boson
($Z'_{SM}$) is considered in the following. With an expected effective 
production cross section $\sigma(e^+e^- \to Z'_{SM})$ of 
$\simeq$~15~pb, including the effects of ISR and luminosity spectrum, a $Z'$ 
resonance will tower over a $q \bar q$ continuum background of 0.13~pb. 
While the observation of such a signal is granted, the accuracy that can be 
reached in the study
of its properties depends on the quality of the accelerator beam energy 
spectrum and on the detector response, accounting for the accelerator induced
backgrounds. One of the main characteristics of the {\sc Clic} collider is the
large luminosity, $L=10^{35}$~cm$^{-2}$~s$^{-1}$ at $\sqrt{s_0}$ = 3~TeV, 
obtained in a regime of strong beamstrahlung effects ($\delta_B=30\%$).
The optimisation of the total luminosity and its fraction in 
the peak has been studied for the case of a resonance scan. The {\sc Clic} 
luminosity spectrum has been obtained with a dedicated beam simulation 
program~\cite{ds} for the nominal parameters at $\sqrt{s_0}$ = 3~TeV. 
In order to study the systematics from the knowledge of this spectrum, 
the modified Yokoya-Chen parametrisation~\cite{peskin} has been adopted. The 
beam energy spectrum is described in terms of $N_\gamma$, the number of 
photons radiated per $e^{\pm}$ in the bunch, the beam energy spread in 
the linac $\sigma_p$ and the fraction $\cal{F}$ of events outside the 
0.5\% of the centre-of-mass energy. 
Two sets of parameters have been considered, obtained by modifying the beam 
size at the interaction point and therefore the total luminosity and its fraction in the highest energy region of the spectrum: CLIC.01 with 
$L$=1.05$\times10^{35}$~cm$^{-2}$ s$^{-1}$ and $N_{\gamma}$=2.2; CLIC.02 with
$L$=0.40$\times10^{35}$~cm$^{-2}$ s$^{-1}$ and $N_{\gamma}$=1.2.

\subsection{$Z'$ Resonance Scan}

The $Z'$ mass and width can be determined by performing either an energy scan,
like the $Z^0$ scan performed at {\sc Lep}/{\sc Slc} and also foreseen
for the $t \bar t$ threshold, or an auto-scan, by tuning the collision energy 
just above the top of the resonance and profiting  of the long tail of the 
luminosity spectrum to probe the resonance peak. For the first method
both di-jet and di-lepton final states can be considered, while for the 
auto-scan only $\mu^+ \mu^-$ final states may provide with the necessary 
accuracy for the $Z'$ energy. $e^+e^- \rightarrow Z'$ events have been 
generated for $M_{Z'}$ = 3~TeV, including the effects of ISR, 
luminosity spectrum and $\gamma \gamma$ backgrounds, assuming SM-like 
couplings, corresponding to a total width $\Gamma_{Z'_{SM}} \simeq$ 90~GeV 
(see Figure~1). 

\begin{figure}[h!]
\begin{center}
\epsfig{file=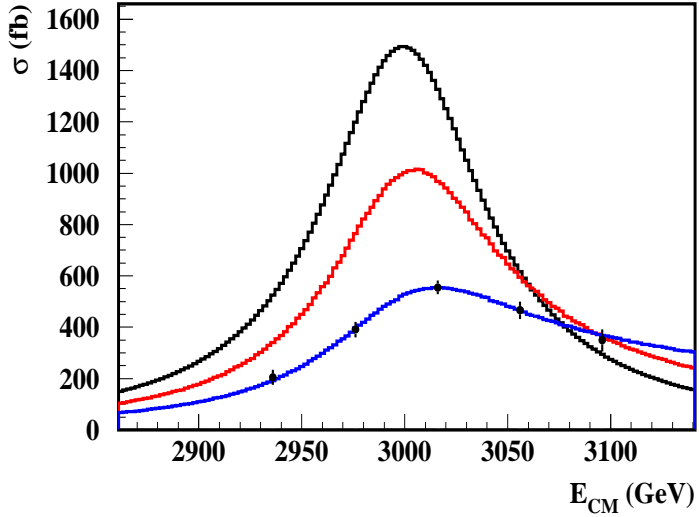,width=11.5cm,height=7.5cm,clip}
\caption[] {The $Z'_{SM} \to \ell^+ \ell^-$ resonance profile obtained by 
energy scan. The Born production cross-section, the cross section with ISR 
included and that accounting for the luminosity spectrum and the tagging 
criteria are shown.}
\label{fig:zpxs}
\end{center} 
\end{figure}
A data set of  1000 fb$^{-1}$ has been assumed for the
CLIC.01 beam parameters and of 400 fb$^{-1}$ for CLIC.02, corresponding to one 
year (10$^{7}$~s) of operation at nominal luminosity. This has been shared
in a 3 to 7 points scan and $M_{Z^{'}}$, $\Gamma(Z^{'})/\Gamma_{SM}$ and 
$\sigma_{peak}$ have been extracted from a $\chi^2$ fit to the predicted
cross section behaviour for different mass and width values. The dilution of 
the analysing power due to the beam energy spread is appreciable, as can be 
seen by comparing the statistical accuracy from a fit to the pure Born cross
section to after including ISR and beamstrahlung effects. Still, the 
relative statistical accuracies are better than 10$^{-4}$ on the mass and 
$5 \times 10^{-3}$ on the width. 
Sources of systematics from the knowledge of the
shape of the luminosity spectrum have also been estimated. In order to keep
$\sigma_{syst} \le \sigma_{stat}$ it is necessary to control 
$N_{\gamma}$ to better than 5\% and the fraction $\cal{F}$ of collisions at 
$\sqrt{s} < 0.995 \sqrt{s_{0}}$ to about 1\%.
\begin{table}[h!]
\caption{Results of the fits for the cross section scan of a $Z'_{SM}$
obtained by assuming no radiation and ISR with the effects of two different
optimisation of the CLIC luminosity spectrum.}
\label{table1}
\begin{tabular}{l c c c}
Observable & Breit Wigner & CLIC.01 & CLIC.02 \\ \hline
$M_{Z^{'}}$ (GeV) & 3000 $\pm$ .12  & $\pm$ .15 &  $\pm$ .21 \\
$\Gamma(Z^{'})/\Gamma_{SM}$ & 1. $\pm$ .001 & $\pm$ .003  & $\pm$ .004 \\
$\sigma^{eff}_{peak}$ (fb) & 1493 $\pm$ 2.0 & 564 $\pm$ 1.7 & 669 $\pm$ 2.9 \\
\hline
\end{tabular}
\end{table}

\subsection{$Z'$ Decays}

The {\sc Clic} potential for the study of the couplings of a new resonance to 
fermions and of its possible decays to new particles has been tested for the 
case of the SM mode $Z' \to c \bar c$, $b \bar b$ and that of the exotic
decay into pairs of right handed Majorana neutrinos. The large boost 
acquired by short-lived heavy hadrons from two fermion production at 3~TeV
provides a very distinctive pattern of detached secondary vertices. A 
tagging technique based on charged multiplicity steps in the vertex tracker, 
independent on the track reconstruction efficiency in the highly collimated 
jets, provides a performance comparable to that achieved at 
{\sc Lep}~\cite{mb}.
Therefore an accuracy on the partial decay widths $\Gamma_{c \bar c}$ and 
$\Gamma_{b \bar b}$ of the order of 10$^{-3}$ can be obtained.

If the observed resonance is the neutral gauge boson $Z'_{LR}$ of the 
Left-Right symmetric model~\cite{lrsm}, the presence of right-handed 
Majorana neutrinos $N_{e}$ opens further decay channels. Their observability
at {\sc Clic} was studied assuming their masses to be generated through the 
See-Saw mechanism~\cite{see-saw}.
Since $N_{e}$ is expected to decay promptly into $e^{\pm}$ and a 
$q_{i}\bar{q}_{j}$ pair, the typical signature consists of two electrons, with
the same charge in half of the cases, and four hadronic jets. Their 
reconstructed energies and momenta need to be recalibrated to correct
for the overlapping $\gamma \gamma \rightarrow \mbox{hadrons}$ events. 
The assignment of the two electrons, $e_{1,2}$, and the four hadronic jets 
$j_{a,b,c,d}$ to their correct $N_{e}N_{e}$ pair is performed by choosing the 
($e_{1}j_{a}j_{b}$\,;\,$e_{2}j_{c}j_{d}$) combination which minimizes the
mass difference $|m(e_{1}j_{a}j_{b})-m(e_{2}j_{c}j_{d})|$. 
The invariant masses of these two selected systems and the resulting $Z'_{LR}$ 
boson mass spectrum are shown in Figure~2.
\begin{figure}[h!]   
\begin{center} 
\epsfig{file=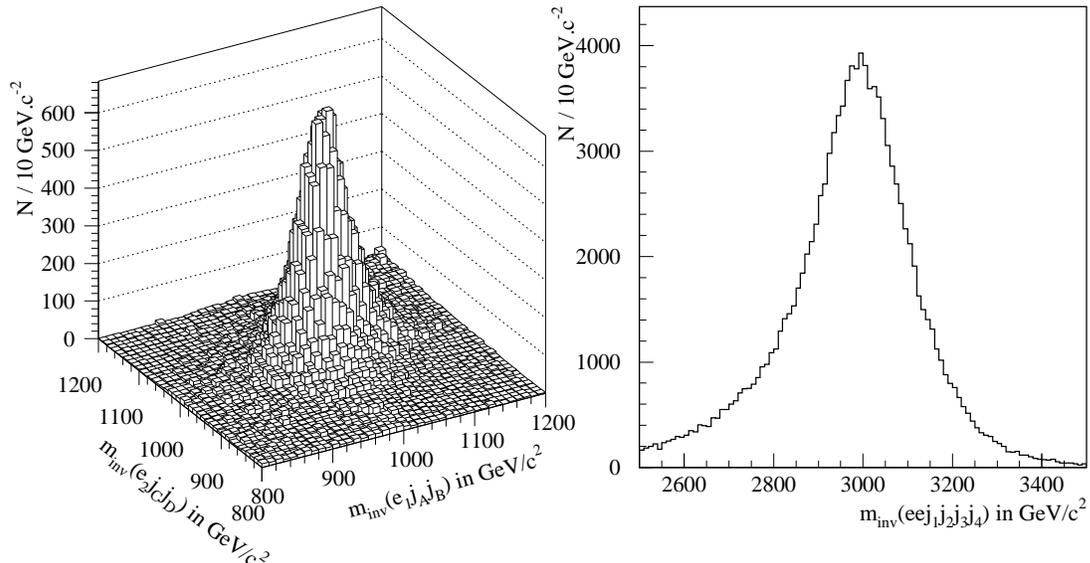,height=7.5cm}
\caption[] {Reconstruction of the right-handed Majorana neutrino $N_{e}$ and 
the $Z'_{LR}$ boson. The spectra correspond to $M_{Z'}$~=~3~TeV and 
$M_{N_{e}}$~=~1~TeV ($L$=1000~fb$^{-1}$).} 
\label{masszrne}
\end{center} 
\end{figure}

\section{Study of the degenerate BESS Model}

While the precise electroweak and {\sc Lep}-2 data favour a realisation of the 
Higgs mechanism through a light elementary Higgs boson, it remains 
important to assess the sensitivity of a future multi-TeV collider to a strong
electroweak symmetry breaking (SSB) scenario. 
SSB models are based on low energy effective lagrangians which provide a 
phenomenological description of the Goldstone boson dynamics. 
Possible new vector resonances produced by the strong interaction responsible 
for the electroweak symmetry breaking can be introduced in the formalism as the
gauge bosons of a hidden symmetry.
A description  of a new triplet of vector resonances is obtained
by considering an effective lagrangian based on the symmetry 
$SU(2)_L\otimes SU(2)_R\otimes SU(2)_{local}$~\cite{bess}. The new vector 
fields are a gauge triplet of the $SU(2)_{local}$.
These new fields acquire mass through the same mechanism which gives mass to 
the $W^{\pm}$ and the $Z^0$ bosons. By enlarging the symmetry group of the 
model, new vector and axial-vector resonances can be described.

The degenerate BESS model (D-BESS)~\cite{dbess} is a realisation of dynamical 
electroweak symmetry breaking with decoupling.  
The D-BESS model introduces two new triplets of gauge bosons, which are 
almost degenerate in mass, ($L^\pm$, $L_3$),
($R^\pm$, $R_3$). The extra parameters  are a new gauge coupling constant
$g''$ and a mass parameter $M$, related to the scale of the
underlying symmetry breaking sector.
In the charged sector the $R^\pm$ fields  are not mixed and $M_{R^\pm}=M$,
while $M_{{L}^\pm}\simeq M (1+x^2)$ where $x=g/g''$ with $g$  the usual 
$SU(2)_W$ gauge coupling constant.
The $L_3$, $R_3$ masses are given by
$M_{L_3}\simeq  M\left(1+ x^2\right),~~ M_{R_3}\simeq
M \left(1+ x^2 \tan^2 \theta\right)$
where $\tan \theta = s_{\theta}/c_{\theta} = g'/g$ and $g'$ is the usual
$U(1)_Y$ gauge coupling constant. These resonances are narrow and almost 
degenerate in mass with $ \Gamma_{L_3}/M\simeq 0.068~ x^2$ and
$\Gamma_{R_3}/M\simeq 0.01~ x^2$, while the neutral mass splitting is:
$\Delta M/M=(M_{L_3}-M_{R_3})/M 
\simeq \left( 1-\tan^2 \theta \right) x^2\simeq 0.70 ~x^2$.
This model respects the existing stringent bounds from electroweak precision 
data since the $S,T,U$ (or $\epsilon_1, \epsilon_2, \epsilon_3$) parameters 
vanish at the leading order due to an additional custodial
symmetry. Therefore, the precision electroweak data only set loose bounds on
the parameter space of the model as shown in Figure~3,
comparable to those from the direct search at the Tevatron~\cite{dbess}.

The behaviour of the resonance widths for this model as well as for
other SM extensions with additional vector bosons is shown in 
Figure~3 as a function of the relevant model parameter. The $Z'_{E_6}$ and 
$Z'_{LR}$ widths are computed by assuming only decays into SM 
fermions and tend to be wider than the corresponding widths of the $L_3$ and
$R_3$ states.

Future hadron colliders may be able to discover these new
resonances which are produced through a $q \bar q$ annihilation and which 
decay in the leptonic channel $q{\bar q'}\to L^\pm,W^\pm\to (e \nu_e)
\mu\nu_\mu$ and $q{\bar q}\to L_3,R_3,Z,\gamma\to 
(e^+e^-)\mu^+\mu^-$.  
\begin{table}[h]
\caption{Sensitivity to $L_3$ and $R_3$ production at the LHC and CLIC 
for $L=$100(500)~fb$^{-1}$ with $M=$1,2(3)~TeV at LHC and
$L=$1000~fb$^{-1}$ at CLIC.}
\begin{tabular}{c c c c c c c}
$g/g''$ & $M$ & $\Gamma_{L_3}$ & $\Gamma_{R_3}$ & $S/\sqrt{S+B}$&
 $S/\sqrt{S+B}$ & $\Delta M$
\\
& (GeV) &(GeV) & (GeV) & LHC ($e+\mu$) & CLIC (hadrons) &  CLIC \\
 \hline  0.1 &
1000 & 0.7 & 0.1 &17.3 & &
\\
0.2 & 1000 & 2.8 & 0.4 & 44.7& &
\\\hline
0.1 & 2000 & 1.4 & 0.2 &3.7& &  
\\
0.2 & 2000 & 5.6 & 0.8 & 8.8& &
 \\\hline
0.1 & 3000 & 2.0 & 0.3 &(3.4)& ~62 & 23.20 $\pm$ .06
\\
0.2 & 3000 & 8.2 & 1.2 &(6.6)& 152 & 83.50 $\pm$ .02
\end{tabular}
\label{dom:table1}
\end{table}  
The relevant observables are the di-lepton transverse and invariant masses.
The main backgrounds, left to these channels after the lepton isolation
cuts, are the Drell-Yan processes with SM gauge bosons
exchange in the electron and muon channel. The study has been performed using 
a parametric detector simulation~\cite{redi}. Results are given in 
Table~\ref{dom:table1} for the combined electron and muon channels
for $L=100$~fb$^{-1}$. Results are given for an integrated luminosity of 
500~fb$^{-1}$ assuming $M=$3~TeV.

The discovery limit at {\sc Lhc} with $L=100$~fb$^{-1}$ is  $M\sim 2$~TeV
with $g/g''=0.1$. Beyond discovery, the possibility to disentangle the double 
peak structure depends strongly on $g/g''$ and smoothly on the 
mass~\cite{redi}.
A lower energy LC can also probe this multi-TeV region through 
the virtual effects in the cross-sections for $e^+e^-\to {L_3},{R_3},Z,\gamma
\to f \bar f $. Due to the presence of new spin-one resonances the 
annihilation channel in $f \bar f$ and $W^+W^-$ is more efficient than
the fusion channel. 
\begin{figure}[hb!]
\begin{center}
\vspace*{-0.75cm} 
\begin{tabular}{ll}
\hspace{-1.0cm}
\mbox{\psfig{file=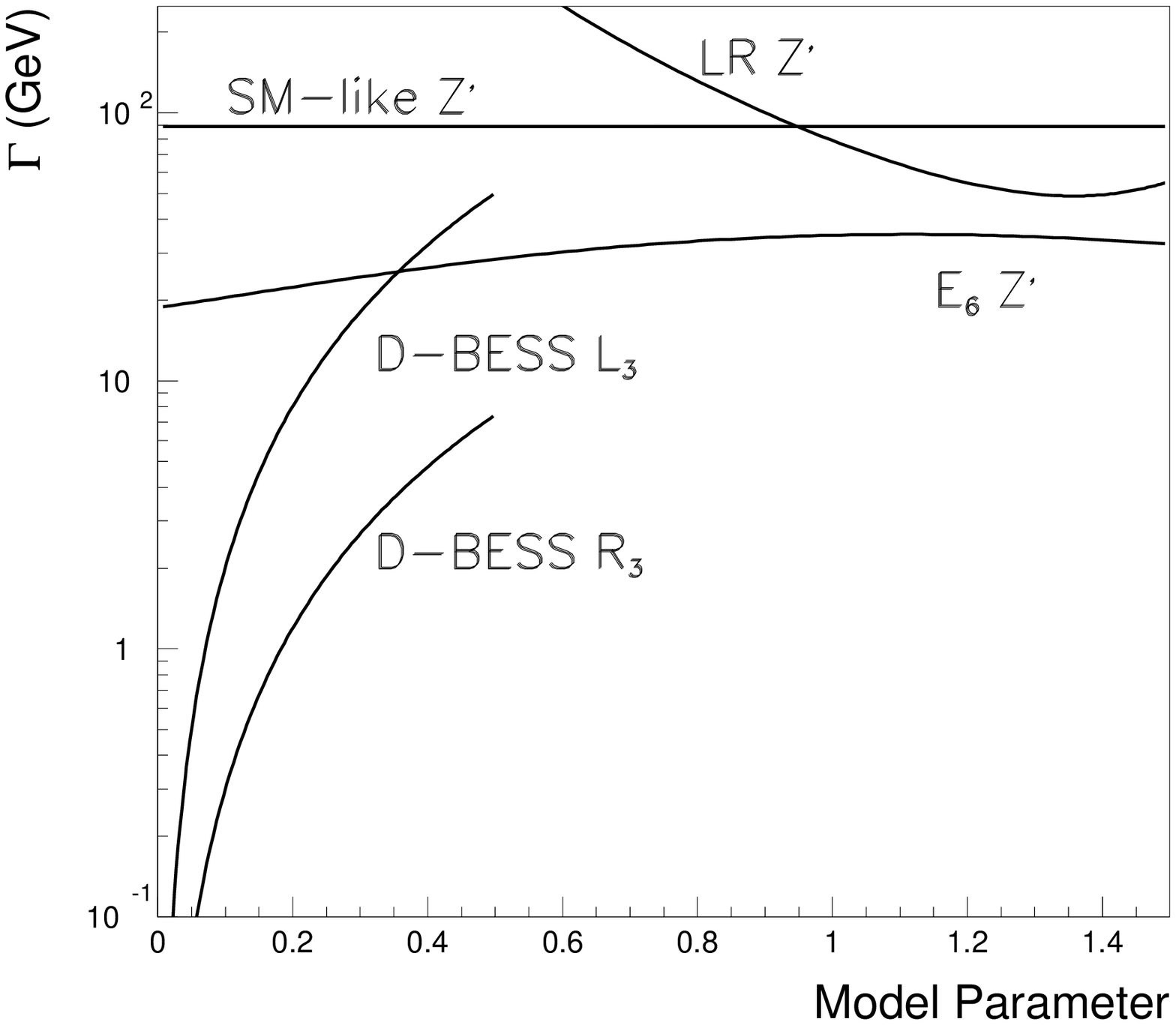,height=6.8truecm}} &
\hspace{-.1cm} 
\mbox{\psfig{file=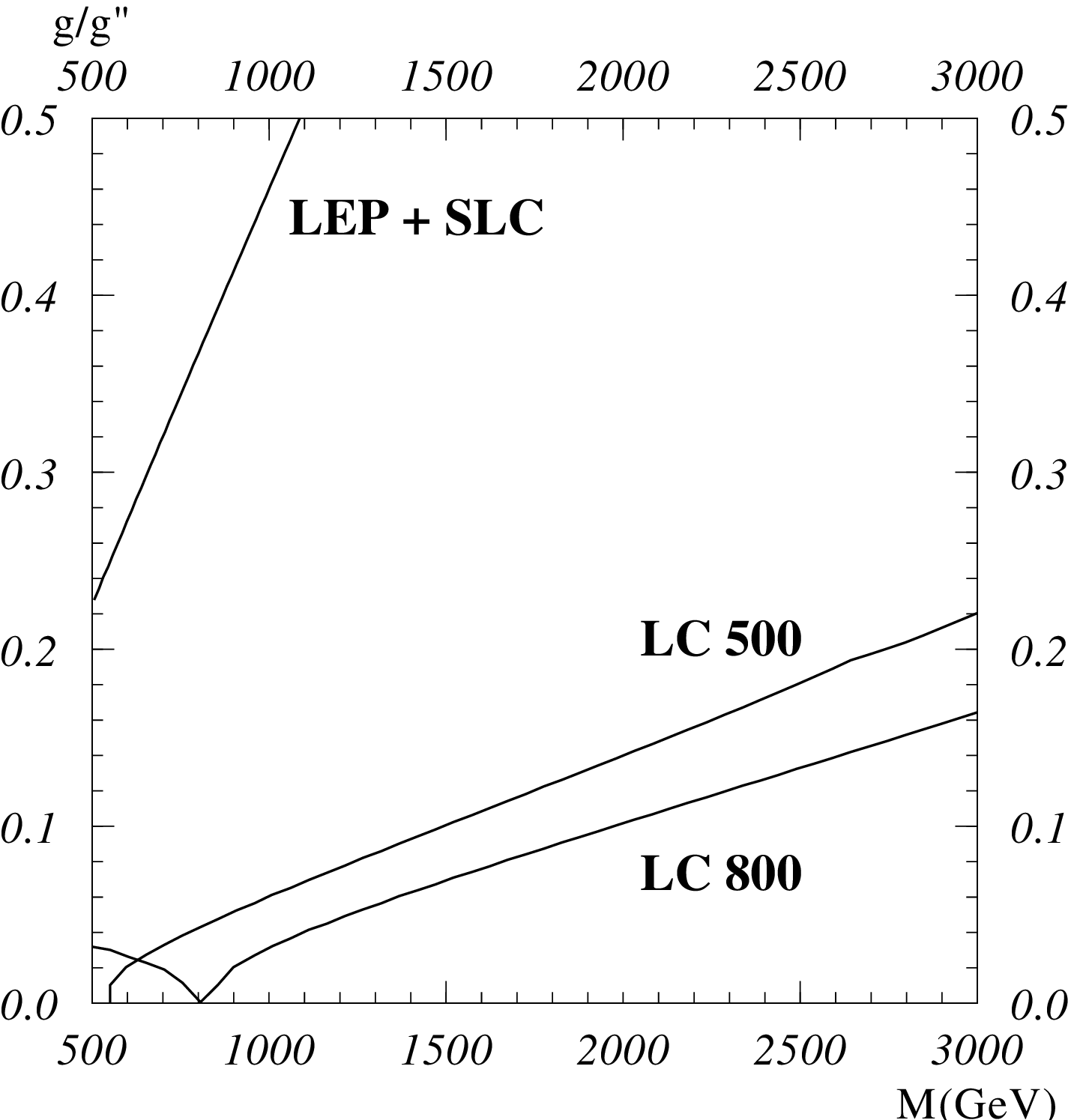,height=6.2truecm}}
\end{tabular}
\caption[]{The left plot shows the widths of the new gauge vectors predicted 
in various new physics models as a function of the relevant parameters: 
$\theta_2$ for $Z'_{E_6}$ and $\lambda=g_L/g_R$ for $Z'_{LR}$~\cite{alta},  
$g/g''$ for D-BESS~\cite{dbess}. The right plot shows the 95\% CL
contour in the plane $(M,~ g/g'')$ from $e^+e^-$ linear
colliders with $\sqrt{s}=500(800)$~GeV. Also shown are the 
present bounds from LEP and SLC. The allowed regions are below the lines.}
\label{dom:fig1}
\end{center}
\end{figure}
In the case of D-BESS, the $L_3$ and $R_3$ states 
are not strongly coupled to $WW$ making the $f\bar f$ final states the most
favourable channel for discovery. The analysis at $\sqrt{s}=$ 500 GeV and 
$\sqrt{s}=$ 800 GeV,is based on the following observables:
$\sigma^{\mu},~~\sigma^h$,  $A_{FB}^{e^+e^- \to \mu^+ \mu^-}$,
$A_{FB}^{e^+e^- \to {\bar b} b}$, $A_{LR}^{e^+e^- \to \mu^+
\mu^-}$, $A_{LR}^{e^+e^- \to {\bar b} b}$, $A_{LR}^{e^+e^- \to
{had}}$.
The sensitivity contours obtained for $L=$ 1000 fb$^{-1}$ and $P(e^-)=80\%$
are shown in Figure~3. The LC indirect reach for $\sqrt{s}<$M is 
lower or comparable to that of the {\sc Lhc}. However, the QCD background 
rejection essential for the {\sc Lhc} sensitivity still needs to be validated 
using full detector simulation and pile-up effects.  

\subsection{Resonance Scan}

Assuming a resonant signal to be seen at the {\sc Lhc} or at a lower energy 
LC, {\sc Clic} can measure its width, mass, and also investigate the existence 
of an almost degenerate structure~\cite{jhep}. 
\begin{figure}[hb!]
\vspace*{-0.75cm} 
\begin{center}
\hspace{-1cm}
\epsfig{file=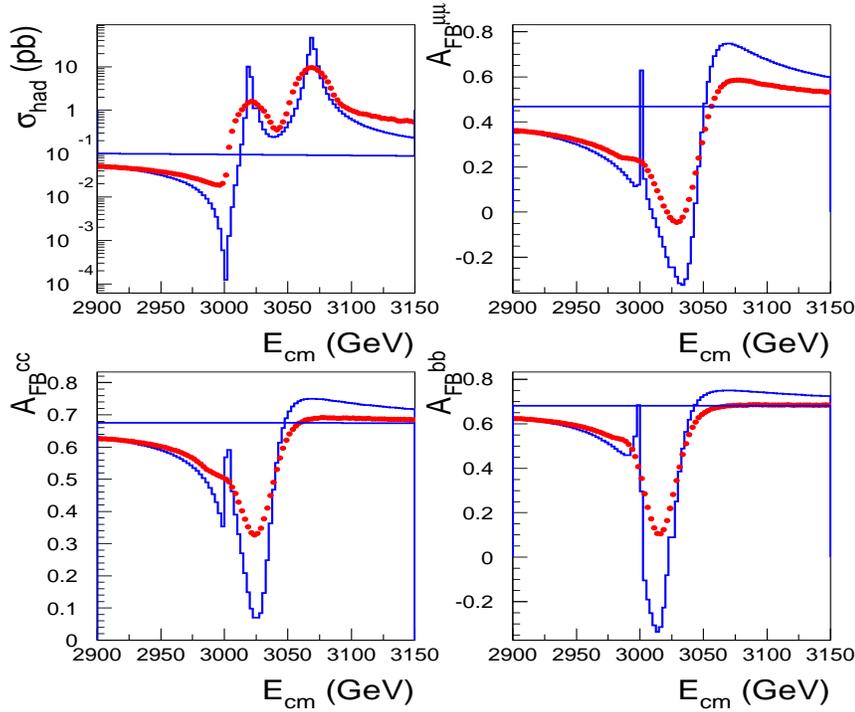,width=13.0cm,height=10.5cm,clip}
\end{center}
\vspace*{-0.25cm} 
\caption[]{The hadronic cross section (upper left) and $\mu^+\mu^-$ (upper 
right), $c \bar c$ (lower left) and $b \bar b$ (lower right) forward-backward
asymmetries at energies around 3~TeV. The continuous lines represent the 
predictions for the D-BESS model with $M$ = 3~TeV and $g/g''=0.15$, the flat 
lines the SM expectation and the dots the observable D-BESS signal after 
accounting for the CLIC.02 luminosity spectrum.}
\label{fig:afb}
\end{figure}
This needs to be validated taking full account of the luminosity
spectrum and accelerator induced backgrounds. The ability in identifying the 
model distinctive features has been studied using the production cross section
and the flavour dependent forward-backward asymmetries, for different values 
of $g/g''$. The resulting distributions are shown in Figure~4 for the case of 
the CLIC.02 beam parameters.
A characteristic feature of the cross section distributions is the presence 
of a narrow dip at energies around 3~TeV. This is due to the interference of  
the $L_3$, $R_3$ resonances with the $\gamma$ and $Z$ and to cancellations of 
the $L_3$, $R_3$ contributions. The effect becomes larger for decreasing  
$g/g''$. Similar considerations hold for the asymmetries. In the case shown
in Figure~\ref{fig:afb}, the effect is still visible after accounting for the
CLIC.02 luminosity spectrum.

This study has demonstrated that with 1000~fb$^{-1}$ of data, {\sc Clic} will 
be able to resolve the two narrow resonances for values of the coupling ratio 
$g/g'' >$~0.08, corresponding to a mass splitting $\Delta M$ = 13~GeV for 
$M=$ 3~TeV, and to determine $\Delta M$ with a statistical accuracy better 
than 100~MeV (see Table~2).   

\section{Conclusions}

Two scenarios predicting new resonances in the multi-TeV region have been 
studied in relations to the discovery and study potential of the {\sc Clic}
$e^+e^-$ linear collider. The profile of such a new resonance can be studied 
with high accuracy due to the large anticipated {\sc Clic} luminosity. This 
accuracy can be exploited to distinguish the nature of the resonance and the 
cases of an additional $Z'$ gauge boson and of nearly degenerate resonances 
from a strong symmetry breaking scenario have been discussed.

\end{document}